# Prediction of $C_7N_6$ and $C_9N_4$: Stable and strong porous carbon-nitride nanosheets with attractive electronic and optical properties

Bohayra Mortazavi[*a], Masoud Shahrokhi[b], Alexander V Shapeev[c], Timon Rabczuk[d] and Xiaoying Zhuang[a]

[a]Institute of Continuum Mechanics, Leibniz Universität Hannover, Appelstraße 11, 30157 Hannover, Germany.
[b]Department of Physics, Faculty of Science, Razi University, Kermanshah, Iran.
[c]Skolkovo Institute of Science and Technology, Skolkovo Innovation Center, Nobel St. 3, Moscow 143026, Russia.
[d]College of Civil Engineering, Department of Geotechnical Engineering, Tongji University, Shanghai, China.

**Abstract**

In this work, three novel porous carbon-nitride nanosheets with $C_7N_6$, $C_9N_4$ and $C_{10}N_3$ stoichiometries are predicted. First-principles simulations were accordingly employed to evaluate stability and explore the mechanical, electronic and optical properties. Phonon dispersions confirm the dynamical stability of all predicted nanosheets. Nonetheless, ab-initio molecular dynamics results indicate that only $C_7N_6$ and $C_9N_4$ are thermally stable. $C_7N_6$, $C_9N_4$ and $C_{10}N_3$ nanosheets were predicted to exhibit high elastic modulus of 212, 202 and 208 N/m and maximum tensile strengths of 14.1, 22.4 and 15.8 N/m, respectively. $C_7N_6$ monolayer was confirmed to be a direct band-gap semiconductor, with a 2.25 eV gap according to the HSE06 method estimation. Interestingly, $C_9N_4$ and $C_{10}N_3$ monolayers show metallic character. The first absorption peaks of optical spectra reveal that $C_7N_6$ nanosheet can absorb the visible light, whereas $C_9N_4$ and $C_{10}N_3$ monolayers can absorb the Infrared range of light. Moreover, the absorption coefficient and optical conductivity of predicted nanosheets in the visible range of light are larger than those of the graphene. The results provided by this study confirm the stability and highlight very promising properties of $C_7N_6$ and $C_9N_4$ nanosheets, which may serve as promising candidates for numerous advanced technologies.

Corresponding author: *bohayra.mortazavi@gmail.com,

# 1. Introduction

Two-dimensional (2D) materials [1,2] are currently among the most attractive family of materials with highly promising application prospects for a wide range of advanced technologies. Graphene as the most prominent member of 2D materials family has been proven to exhibit exceptionally high mechanical [3] and thermal conduction [4] properties along with unique optical and electronic features [5–9]. In general, for many applications in nanoelectronics and nanooptics presenting a narrow and direct band-gap is among the most important requirements, however graphene is a zero band-gap semimetal. Worthy to remind via chemical doping with nitrogen or boron atoms [10–12], mechanical straining [13–15] or creation of nanomesh structures [16,17], a band-gap can be opened in the graphene. Nevertheless, these aforementioned methods require additional processing steps after the first synthesize of graphene, which question their practical application due to the increase in the fabrication time and cost as well. Such that designing novel 2D materials that are semiconductors in their pristine form has been among the most appealing approaches for the employment of 2D materials in nanoelectronics and nanophotonics.

Among the various classes of 2D materials, carbon-nitride nanomembranes have been among the most successful nanomaterials with inherent semiconducting electronic characters. These 2D allotropes show a common chemical formula $C_xN_y$, where x and y represent the number of C and N atoms in the unit cell, respectively. Carbon-nitride nanomaterials offer stiff and stable components owing to the formation of strong covalent bonds between the C-C and C-N bonds. Depending on the composition of C and N atoms in various atomic lattices, carbon-nitride allotropes can exhibit diverse electronic, optical, electrochemical, mechanical and thermal conduction properties. From few decades ago, graphitic carbon nitride layered materials with a chemical formula of $C_3N_4$, have been synthesized via polymerization of cyanamide, dicyandiamide or melamine [18]. Unlike the graphite, graphitic carbon nitrides show porous atomic lattices and low thermal conductivities [19] and more importantly are inherent semiconductors. Graphitic carbon nitrides show good mechanical properties and have been proven as promising candidates for the employment; in energy conversion/storage, catalysis, photocatalysis and oxygen reduction systems [18,20–23]. A decade after the first experimental realization of graphene by the mechanical exfoliation [1], Siller *et al.* [24] reported the first successful synthesis of $C_3N_4$ graphitic carbon-nitride nanosheets employing an ionothermal interfacial reaction. Nitrogenated holey graphene with a $C_2N$ stoichiometry in the 2D form was experimentally fabricated by Mahmood *et al.* [25], via a wet-chemical reaction. $C_3N_4$ and $C_2N$ nanosheets were found to be semiconductors with around two orders of magnitude lower thermal conductivities than that of the pristine graphene [19]. Mahmood *et al.* [26], reported the first experimental realization of a densely packed and graphene-like carbon-nitride nanomaterials with a $C_3N$ stoichiometry. Likely to $g-C_3N_4$ and $C_2N$ counterparts, $C_3N$ is also an intrinsic semiconductor, however thanks to its nonporous atomic lattice $C_3N$ can exhibit remarkably high mechanical and thermal conduction properties [27–32]. According to the extensive theoretical studies conducted recently, $C_3N$ nanosheet can exhibit desirable properties for various applications, like; nanoelectronics [33–37], nanomagnetics [38,39], catalysis [40,41], superconductivity [42], anode materials for Li-ion batteries [43] and hydrogen storage [44].

Recent advances with respect to the fabrication of carbon-nitride nanomaterials have undoubtedly enhanced the prospect for the experimental realization of other novel nanosheets made merely from covalent networks of carbon and nitrogen atoms. In addition, the comprehensive insight provided by the extensive theoretical studies have proven that carbon-nitride nanosheets can serve as promising candidates for a wide range of applications and in some cases can outperform the graphene. On this basis, a simple but important question is that if other carbon-nitride nanosheets can stay stable under the ambient conditions, and if yes, what about their intrinsic electronic, mechanical, optical and thermal properties. To address these questions, state-of-the-art theoretical studies can be considered as the feasted but also the most practical approaches in order to examine the stability of new compositions, estimate their properties, find potential applications and suggest possible synthesis routes [45–48]. Such that, experimental fabrication endeavors for a novel material can be initiated and encouraged by the theoretical confirmations of the stability and promising intrinsic properties. In this work motivated by the exceptional properties of carbon-nitride nanosheets, we predicted three novel porous lattices; $C_7N_6$, $C_9N_4$ and $C_{10}N_3$. Stability, mechanical properties and electronic/optical properties of these novel nanomembranes were explored by the first-principles density functional theory simulations. Acquired theoretical results confirm the stability and highlight highly attractive properties of $C_7N_6$ and $C_9N_4$ nanosheets, and may hopefully motivate and guide future experimental and theoretical studies.

## 2. Computational methods

Density functional theory (DFT) calculations in this work were performed using the *Vienna Ab-initio Simulation Package* (VASP) [49–51] within the generalized gradient approximation (GGA) and Perdew−Burke−Ernzerhof (PBE) [52], considering a plane-wave cutoff energy of 500 eV. The convergence criterion for the electronic self-consistent-loop was set to $10^{-4}$ eV.

Periodic boundary conditions were applied along all three Cartesian directions, with a vacuum layer of 15 Å to avoid the interactions along the normal direction of the monolayers. Uniaxial tensile simulations were performed to evaluate the mechanical properties. After applying the changes in the simulation box size, conjugate gradient method was used for the rearrangements of atomic positions and reaching to the energy minimized structure. In this case, the convergence criteria for the forces on each atom was taken to be 0.01 eV/Å, in which we defined a 5×5×1 Monkhorst-Pack [53] k-point mesh size. The electronic properties were evaluated using a denser k-point grid of 15×15×1 within the tetrahedron method with Blöchl corrections [54]. The screened hybrid functional HSE06 [55] was employed particularly to provide more accurate estimations for the band-gap value. To evaluate the force constants, density functional perturbation theory (DFPT) simulations within the local-density approximation (LDA) were conducted over 2×2×1 super-cell structures. Phonon dispersions were acquired using the PHONOPY code [56], with the input provided by the DFPT calculations. Thermal stability was examined by performing the ab-initio molecular dynamics (AIMD) simulations for 20 ps with a time step of 1 fs, over 2×2×1 super-cells, in which a 2×2×1 k-point mesh size was applied.

Optical calculations were performed in the random phase approximation (RPA) over the PBE results using the all electron full potential as implemented in the Wien2k [57] code. In order to achieve energy eigenvalues convergence, wave functional in the interstitial region were expanded in terms of plane waves with a cut-off parameter of RMT×$K_{max}$=8.5, where RMT denotes the smallest atomic sphere radius and $K_{max}$ largest *k* vector in the plane wave expansion and the maximum angular momentum of the atomic orbital basis functions was set to $l_{max}$=10. The plane-wave cutoff was set to $G_{max}$=12 and a 24×24×1 Γ centered k-point mesh was used to calculate the optical spectra. The imaginary part of the dielectric function by considering interband transitions is obtained using of following relation:

$$\mathrm{Im}\,\varepsilon_{\alpha\beta}^{(\mathrm{inter})}(\omega) = \frac{4\pi^2 e^2}{\Omega} \lim_{q \to 0} \frac{1}{|q|^2} \sum_{c,v,k} 2w_k \delta(\varepsilon_{ck} - \varepsilon_{vk} - \omega) \times \langle u_{ck+e_\alpha q} | u_{vk} \rangle \langle u_{ck+e_\beta q} | u_{vk} \rangle^* \quad (1)$$

where $q$ is the Bloch vector of the incident wave, $w_k$ is the **k**-point weight and the band indices *c* and *v* are restricted to the conduction and the valence band states, respectively. The real part of dielectric function can be evaluated from $\mathrm{Im}\,\varepsilon_{\alpha\beta}(\omega)$ using the Kramers–Kronig transformation:

$$\mathrm{Re}\,\varepsilon_{\alpha\beta}^{(\mathrm{inter})}(\omega) = 1 + \frac{2}{\pi} P \int_0^\infty \frac{\omega' \mathrm{Im}\,\varepsilon_{\alpha\beta}(\omega')}{(\omega')^2 - \omega^2 + i\eta} d\omega' \quad (2)$$

where $P$ denotes the principle value and $\eta$ is the complex shift. In the metallic systems ($C_9N_4$ and $C_{10}N_{13}$ monolayers), the intraband transitions contribution was also included to the interband transitions to investigate the optical properties [58,59]:

$$\operatorname{Im}\varepsilon_{\alpha\beta}^{[intra]}(\omega) = \frac{\Gamma\omega_{pl,\alpha\beta}^2}{\omega(\omega^2+\Gamma^2)} \quad (3)$$

$$\operatorname{Re}\varepsilon_{\alpha\beta}^{[intra]}(\omega) = 1 - \frac{\omega_{pl,\alpha\beta}^2}{\omega(\omega^2+\Gamma^2)} \quad (4)$$

where $\omega_{pl}$ is the plasma frequency and $\Gamma$ is the lifetime broadening.

## 3. Results and discussions

Energy minimized $C_7N_6$, $C_9N_4$ and $C_{10}N_3$ monolayers predicted in this study are shown in Fig. 1, which show hexagonal atomic lattices. These novel nanosheets are made of three pentagon cores that are connected by single N atoms, forming porous structures with repeating 12 and 9 membered rings made from covalent networks of C and N atoms. The energy per atoms for $C_7N_6$, $C_9N_4$ and $C_{10}N_3$ monolayers was calculated to be -8.418, -8.464 and -8.436 eV, respectively, which are very close. The hexagonal lattice constant of $C_7N_6$, $C_9N_4$ and $C_{10}N_3$ monolayers was measured to be 6.794, 6.875 and 6.948 Å, respectively, which are considerably close in accordance with our earlier observation for lattice energies. However, as it is clear by increasing the C atoms content the lattice constant very slightly increases ~2%. Notably, the C-N bond lengths that connect the pentagon cores, were found to be convincingly the same in the all considered monolayers, ~1.289 Å (maximum difference of less than 0.1%). In the single-layer $C_7N_6$, $C_9N_4$ and $C_{10}N_3$ the C-C bond length in the 9 membered rings was calculated to be 1.478, 1.557 and 1.607 Å, respectively. The C-C bond length in the 12 membered rings of $C_9N_4$ and $C_{10}N_3$ was found to be very close and ~1.452 Å, which is slightly larger than the corresponding C-N bond in the $C_7N_6$, 1.464 Å. As it is clear, the C-N bonds that act as the connector of pentagon cores, show the smallest bond lengths, implying their high rigidity. In contrast the C-C bonds in the 9 membered rings are the longest bonds in these nanosheets. In order to facilitate the future studies, the stress-free energy minimized monolayers are provided in the supplementary information document. Although the carbon-nitride nanosheets are known to be made of covalent bonds, to further confirm this concept the electron localization function (ELF) [60] within the unit-cells is also illustrated in Fig. 1. ELF is a spatial function and takes a value between 0 and 1. It is clear that the electron localization occurs around the center of all the C-C and C-N bonds, and thus confirming the dominance of covalent bonding in these novel nanomaterials. Strong electron localization is also apparent around the N atoms that act as connectors of pentagon cores.

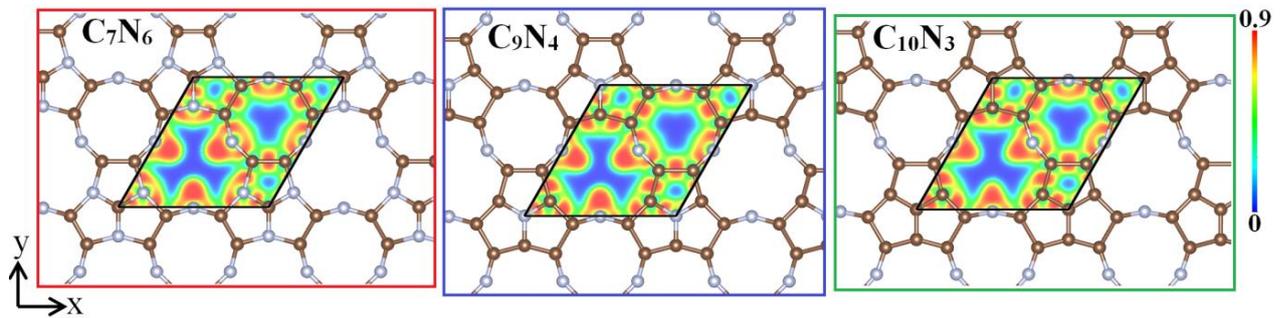

**Fig. 1**, Atomic structure of $C_7N_6$, $C_9N_4$ and $C_{10}N_3$ monolayers. Contours illustrate the electron localization function [60] within the unit-cell.

As a common approach with respect the analysis of the stability of 2D materials, we first examine the dynamical stability of the predicted nanosheets by calculating the phonon dispersion relations along the high symmetry directions of the first Brillouin zone, and the acquired results are shown in Fig. 2. Notably, for the all predicted carbon-nitride monolayers, the dynamical matrix is free of imaginary eigenvalues, which would appear in the phonon dispersion as negative frequencies, and thus confirming the dynamical and structural stability of these novel structures. As expected as the signature of 2D materials, in the all cases two of the three acoustic modes present linear dispersion, while the remaining one exhibits a quadratic dispersion. The dynamical stability also suggest that in the future studies the thermal conductivity of these novel nanosheets can be evaluated by either first-principles Boltzmann transport method or classical molecular dynamics simulations [61–64]. Thermal stability is another critical factor in determining the stability and reducibility of a novel material. To assess the thermal stability, for the all predicted nanosheets the AIMD simulations were conducted at 300, 500 and 1000 K for 20 ps long simulations. Our results reveal that $C_7N_6$ nanosheet can stay completely intact at the all studied temperatures (see Fig. S1a). $C_9N_4$ shows perfect stability at 500 K (see Fig. S1b) but disintegrate at 1000 K. Nevertheless, the $C_{10}N_3$ could not endure at any of the considered temperatures and thus its fabrication is unlikely. The dynamical and thermal stability of $C_7N_6$ and $C_9N_4$ nanosheets are important findings and may serve as encouraging observations toward their experimental realization. Nonetheless, despite of the thermal instability of $C_{10}N_3$ monolayer, we also study its mechanical, electronic and optical properties in order to provide a more comprehensive vision concerning these novel 2D systems.

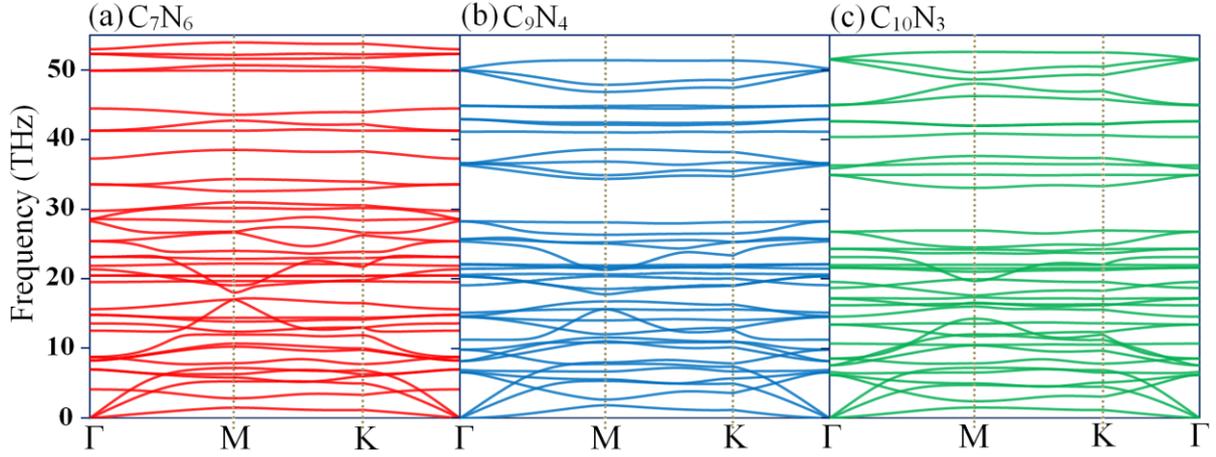

**Fig. 2**, Phonon frequencies of $C_7N_6$, $C_9N_4$ and $C_{10}N_3$ monolayers along the high symmetry directions of the first Brillouin zone.

We next investigate the mechanical properties of $C_7N_6$, $C_9N_4$ and $C_{10}N_3$ nanosheets on the basis of uniaxial tensile simulations. In order to examine the anisotropy of the mechanical properties, the tensile responses were studied by performing the uniaxial loading along the *x* and *y* directions, as depicted in Fig. 1. We remind that for the uniaxial tensile simulations, the periodic simulation box size along the loading direction was increased gradually with a fixed strain step. After applying the loading strain, the simulation box size along the sheet perpendicular direction of the loading was adjusted to reach a negligible stress (<0.02 N/m). The DFT predictions for the uniaxial stress-strain responses of $C_7N_6$, $C_9N_4$ and $C_{10}N_3$ monolayers elongated along *x* and *y* directions are illustrated in Fig. 3. Likely to other conventional materials, the stress-strain curves exhibit initial linear responses, corresponding to the linear elasticity. The acquired results indicate that for the all predicted nanosheets, the initial linear parts of the stress-strain relations coincide very closely for the both considered loading directions, thus confirming the isotropic elasticity. Interestingly, it is clear from the results shown in Fig. 3 that the initial linear responses of $C_7N_6$, $C_9N_4$ and $C_{10}N_3$ are also considerably close. According to our results, the elastic modulus of $C_7N_6$, $C_9N_4$ and $C_{10}N_3$ nanosheets were found to be 212, 202 and 208 N/m, respectively. These results show that the elastic modulus is not very sensitive to the C atoms content in the predicted nanosheets. Within the elastic range, the strain along the traverse direction of loading ($s_t$) with respect to the loading strain ($s_l$) is constant and can be used to evaluate the Poisson's ratio, using: $-s_t/s_l$ [65–67]. On this basis the Poisson's ratio of $C_7N_6$, $C_9N_4$ and $C_{10}N_3$ monolayers was estimated to be 0.24, 0.25 and 0.26, respectively.

On the other hand, according to the acquired stress-strain relations the predicted nanosheets exhibit higher tensile strengths and strain at tensile strength along *x* than *y* direction. This observation indicates anisotropic tensile responses of the studied carbon-nitride films. Moreover, after the tensile strength point, the stress values in the all predicted nanosheets drop sharply, which might be an indication of brittle failure mechanism in these structures. Notably, the mechanical properties of $C_7N_6$ and $C_{10}N_3$ are very close, and distinctly lower than those of the $C_9N_4$ counterpart. Interestingly, $C_9N_4$ can exhibit considerably high tensile strengths of 22.4 and 19.5 N/m at corresponding strain values of 0.14 and 0.12 for the uniaxial loading along the *x* than *y* directions, respectively. Surprisingly, $C_7N_6$ nanosheet which shows the highest thermal stability among the predicted nanomaterials yield the lowest tensile strengths of 14.1 and 11.9 N/m at corresponding strain values of 0.09 and 0.07 along the *x* than *y* directions, respectively. For the case of $C_{10}N_3$ monolayer, despite of its thermal instability it can yet show high tensile strengths of 15.8 and 13.2 N/m, when uniaxially loaded along the *x* than *y* directions, respectively. The elastic modulus (*E*) and maximum tensile strength (*MTS*) of pristine graphene, were reported to be 350.7 and 40.4 N/m by Liu *et al*. [68]. $C_3N$ yields an elastic modulus and maximum tensile strength of around 341.4 and 35.2 N/m [69], which are very close to those of pristine graphene. As expected, porous atomic lattices of predicted nanosheets resulted in exhibiting comparable but distinctly lower mechanical properties than graphene and $C_3N$ counterparts, with densely packed atomic lattices. However, when comparing with other well-known porous carbon-nitride lattices, like tri-triazine-based graphitic carbon-nitride (with *E*=165 N/m and *MTS*=9.95 N/m [70]) and $C_2N$ (with *E*=140 N/m and *MTS*=16.26 N/m [70]), the predicted nanosheets are more rigid and stronger as well. Notably, $C_7N_6$ and $C_{10}N_3$ exhibit very close mechanical properties to the s-triazine-based graphitic carbon-nitride (with *E*=212 N/m and *MTS*=15.22 N/m [70]). As it is clear, the $C_9N_4$ shows the highest tensile strengths among the all well-known carbon-nitride nanosheets with the porous atomic lattices. Notably, the mechanical properties of predicted nanosheets are higher than some of other densely packed 2D materials, like; silicene, germanene and stanene [71].

In Fig. S2, we studied the failure mechanism of predicted nanosheets for the uniaxial loading along the *x* and *y* directions. As a general observation for the uniaxial loading, the bonds oriented along the loading direction stretches and contribute to the load bearing, whereas the bonds oriented along the perpendicular direction of loading remain less affected. The acquired results shown in Fig. S2 reveal that for the studied nanosheets along the different loading directions, first rupture always occurs along a bond that is included in the outer boundary of the three pentagon cores. It can be thus concluded that the in the case of $C_7N_6$ the C-N bonds of pentagon cores and for the $C_9N_4$ and $C_{10}N_3$ nanosheets the corresponding C-C bonds are the weakest bonds in these systems.

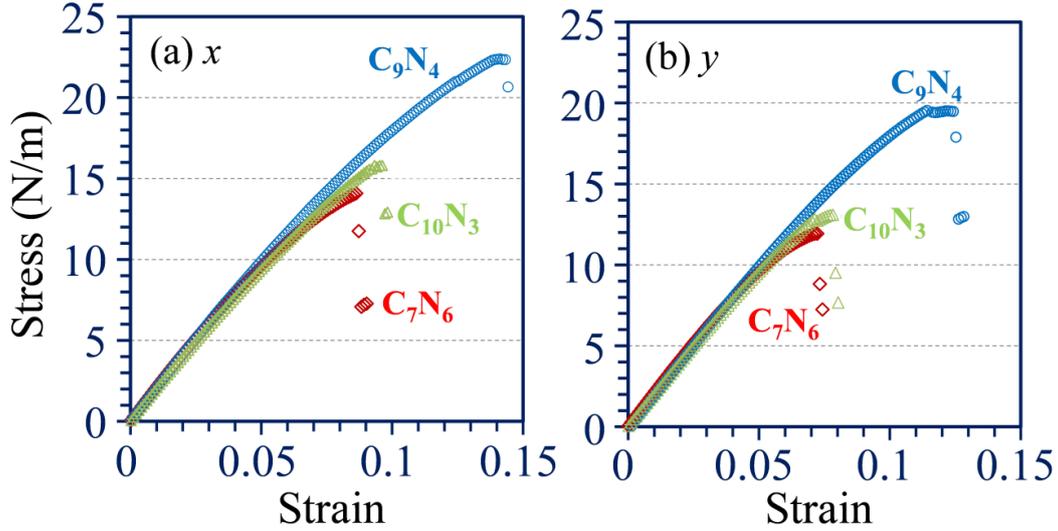

**Fig. 3**, Uniaxial stress-strain responses of $C_7N_6$, $C_9N_4$ and $C_{10}N_3$ nanosheets elongated along the (a) *x* and (b) *y* directions, as depicted in Fig. 1.

The first-principles results presented up to this stage confirm the thermal, dynamical and elastic stability of $C_7N_6$ and $C_9N_4$ nanosheets and also highlight that these novel 2D systems can endure under severe loading conditions, owing to their strong covalent bonding nature. Nonetheless, these observations cannot confirm that the predicted lattices are the global minimum structures. To address this important issue, an extensive search and prediction of various possible structures should be conducted, which requires the employment of crystal structure prediction methods, such as the active learning and evolutionary techniques [45,72]. We would like to also note that he predicted porous atomic lattices in this work can be also used as a template to explore other stable 2D compositions.

We next explore the electronic/optical characteristics of the predicted nanosheets. To study electronic properties of $C_7N_6$, $C_9N_4$ and $C_{10}N_3$ monolayers, we calculate the electronic band structures along the high symmetry *Γ-M-K-Γ* directions using the PBE/GGA and HSE06 methods. Obtained results illustrated in Fig. 4 reveal that the valence band maximum (VBM) and the conduction band minimum (CBM) in $C_7N_6$ monolayer occur at the *Γ*-point in the Brillouin zone, confirming a direct band-gap semiconducting electronic character. According to the PBE and HSE06 results, the band-gaps of $C_7N_6$ monolayer was predicted to be 1.62 and 2.25 eV, respectively. Our analysis of partial electronic density of states (DOS) reveals that in the case of $C_7N_6$ monolayer, the $p_z$ orbitals of the C and N atoms are responsible for the band-gap opening. For the $C_9N_4$ and $C_{10}N_3$ nanosheets, the Fermi level is however crossed by both $\sigma$ ($s+p_x+p_y$) and $\pi$ ($p_z$) orbitals of N and C atoms, leading to the formation of metallic character. According to the band structure predicted by the both PBE and HSE06 methods for the $C_9N_4$ and $C_{10}N_3$ monolayers, the conduction and valence bands overlap at Fermi level, demonstrating metallic electronic signature (results for total DOSs are shown in Fig. S3). Observation of metallic electronic nature in $C_9N_4$ and $C_{10}N_3$ nanosheets is anomalous finding, because most of the carbon-nitride 2D systems are semiconductors.

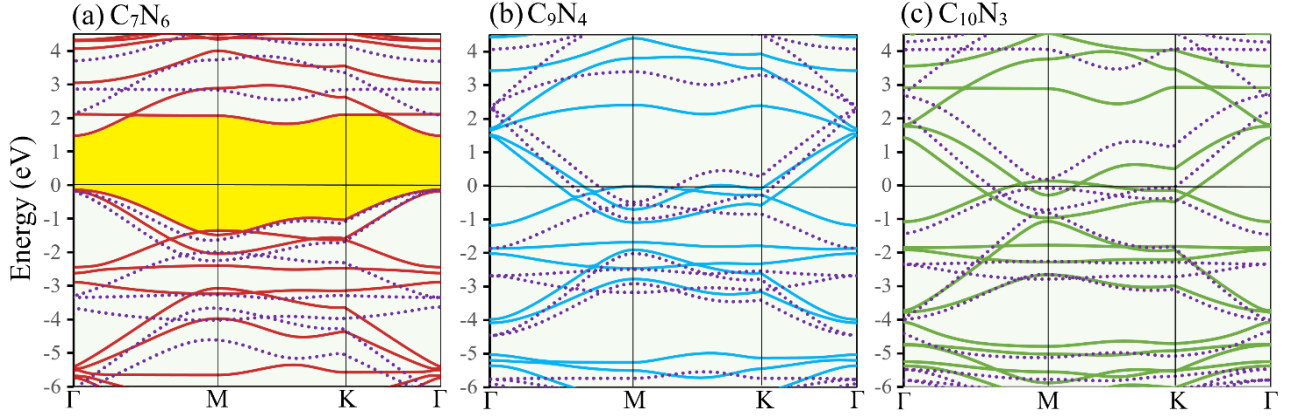

**Fig. 4**, Band structures of single-layer $C_7N_6$, $C_9N_4$ and $C_{10}N_3$ predicted by the PBE/GGA (continuous lines) and HSE06 (dotted lines) functionals. The Fermi energy is aligned to zero.

We next shift our attention to the optical properties of single-layer $C_7N_6$, $C_9N_4$ and $C_{10}N_3$. The optical properties of these monolayers are explored using the RPA [73] method by taking into account the contribution of inter- and intra-band transitions for metallic cases. It is found that the optical spectra become anisotropic along the in-plane and out-of-plane directions. The imaginary and real parts of the dielectric function (Im ($\varepsilon_{\alpha\beta}$) and Re ($\varepsilon_{\alpha\beta}$), respectively) of $C_7N_6$, $C_9N_4$ and $C_{10}N_3$ monolayers versus photon energy for the in-plane and out-of-plane polarization directions obtained from RPA+PBE are illustrated in Fig. 5. The absorption edge of Im ($\varepsilon_{\alpha\beta}$), which is equal to optical-gap, for $C_7N_6$ monolayer occurs at 1.60 and 2.98 eV for the in-plane and out-of-plane polarization. This optical gap further indicates that $C_7N_6$ monolayer is a semiconductor. The first main peaks of Im ($\varepsilon_{\alpha\beta}$) along the in-plane polarization for this monolayer happen in the visible range of light (~2 eV) which are related to $\pi \rightarrow \pi^*$ transitions. The value of the static dielectric constant (the real part of the dielectric constant at zero energy) for $C_7N_6$ monolayer was measured to be 3.58 along the in-plane polarization and 1.38 along the out-of-plane polarization, respectively.

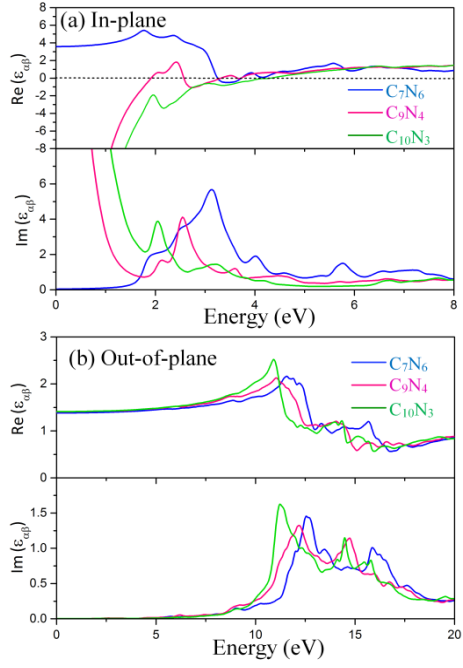

**Fig. 5,** Imaginary and real parts of the dielectric function for the single-layer $C_7N_6$, $C_9N_4$ and $C_{10}N_3$ for the in-plane and out-of-plane polarizations, predicted using the RPA+PBE approach.

In contrast, by taking into account the intraband transition contribution for $C_9N_4$ and $C_{10}N_3$ systems, the imaginary and real parts of the dielectric function show singularity at zero frequency along the in-plane direction, presenting metallic optical responses. It is known that the roots of Re ($\varepsilon_{\alpha\beta}$) with x = 0 line show the plasma frequencies, such that the first root was calculated to be 3.53 and 4.03 eV for $C_9N_4$ and $C_{10}N_3$ monolayers. While, the positive values of the static dielectric constant and nearly zero first plasma frequency confirm the semiconducting characters for the optical spectra along the out-of-plane direction. This can be attributed to the huge depolarization effect that occurs along the out-of-plane direction in 2D materials. These properties have been also reported for other metallic and semi-metallic 2D systems, like; $B_2C$ [74], $Mo_2C$ [59], MXene monolayers [75] and graphene [76,77]. The value of the static dielectric constant for the $C_9N_4$ and $C_{10}N_3$ monolayers were measured to be 1.40 and 1.42 along the out-of-plane polarization, respectively. The main peaks along the out-of-plane direction for all nanosheets are broad and occur at energy range between 10.00 and 12.50 eV, which are related to $\pi \to \sigma^*$ and $\sigma \to \pi^*$ transitions.

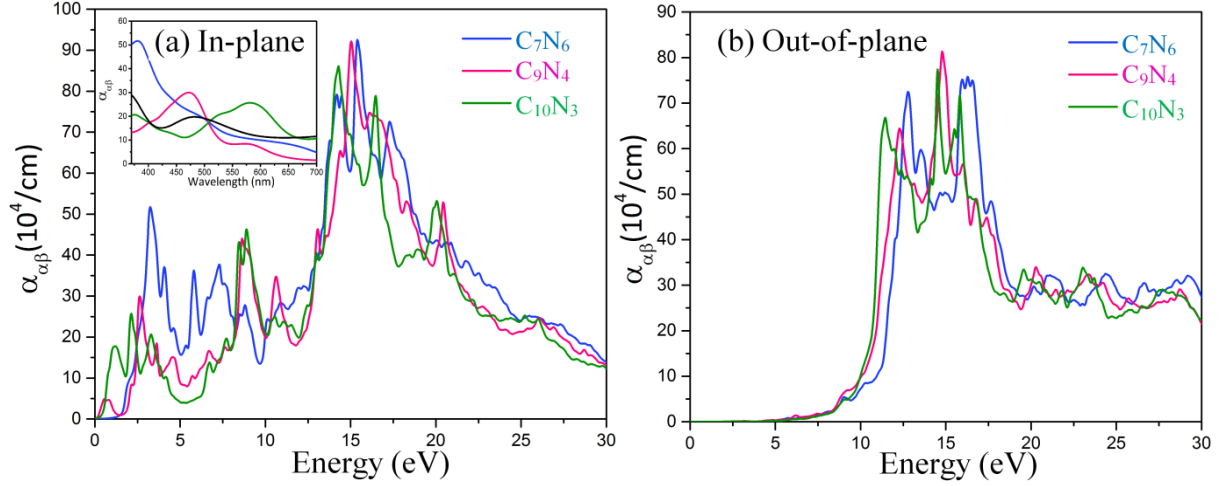

**Fig. 6,** Optical absorption spectra, $\alpha_{\alpha\beta}$, of single-layer $C_7N_6$, $C_9N_4$ and $C_{10}N_3$ as a function of photon energy for the in-plane and out-of-plane polarization, calculated using the RPA+PBE approach. Inset shows a comparison of optical absorption spectra for the in-plane polarization as a function of wavelength, for aforementioned nanosheets and graphene [77,78] (black line) in the visible range (370–700 nm) of light.

In the next step, we discuss the absorption coefficient, $\alpha_{\alpha\beta}(\omega)$, which can be calculated with the following equation [58];

$$\alpha_{\alpha\beta}(\omega) = \frac{\omega \operatorname{Im}\varepsilon_{\alpha\beta}(\omega)}{cn_{\alpha\beta}(\omega)} \qquad (5)$$

where $n_{\alpha\beta}(\omega)$ is the refraction index. The absorption coefficients for the $C_7N_6$, $C_9N_4$ and $C_{10}N_3$ monolayers are plotted in Fig. 6. To provide a better vision, in this case we also compare the absorption coefficient along in-plane polarization of these structures with our previous theoretical results for graphene [77,78] in the visible range of light (from 370 to 700 nm) as a function of wavelength (see Fig. 6 inset). The obtained results indicate that the first absorption peak for the single-layer $C_7N_6$ structure is 1.87 eV along the in-plane polarization, which is in the visible range of light, whereas the corresponding value for the $C_9N_4$ and $C_{10}N_3$ monolayers is 0.52 and 0.77 eV, respectively, which are in the Infrared (IR) range of light. The main peaks of $\alpha_{\alpha\beta}(\omega)$ of $C_7N_6$, $C_9N_4$ and $C_{10}N_3$ monolayers for the in-plane polarization were found to be broad and locate at an energy range between 14.20 and 15.40 eV. The first absorption peak for $C_7N_6$, $C_9N_4$ and $C_{10}N_3$ monolayers along the out-of-plane polarization occurs at 5.77, 4.45 and 4,93 eV, respectively, which are in the ultraviolet (UV) spectral range of light, while the main absorption peaks in this direction locate at energy levels between 14.50 and 16.60 eV. From the results shown in Fig. 6 inset it can be concluded that the absorption coefficients for the $C_7N_6$ and $C_9N_4$ monolayers at wavelength range between 370-500 nm range (from violet to green in visible spectrum) are larger than those of the graphene (the black line in Fig. 6 inset). On the other hand, for the $C_{10}N_3$ monolayer only at wavelength ranges between 500-650 nm (over the yellow range), the optical conductivity is larger than the graphene one. These results highlight that $C_7N_6$ and $C_9N_4$ nanosheets can enhance visible-light absorption in comparison with the graphene, which can be potentially attractive for the photovoltaics applications.

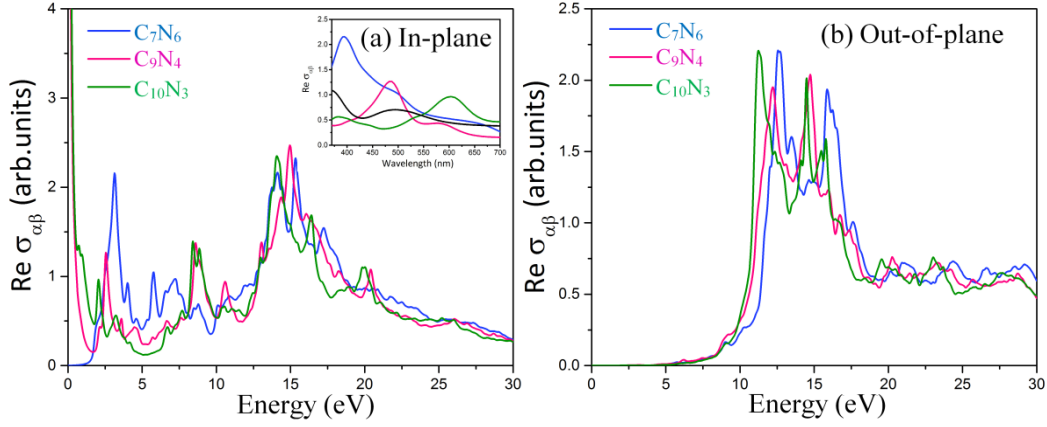

**Fig. 7,** Optical conductivity, Re $\sigma_{\alpha\beta}$, of single-layer $C_7N_6$, $C_9N_4$ and $C_{10}N_3$ as a function of photon energy for the in-plane and out-of-plane polarization, calculated using the RPA+PBE approach. Inset shows a comparison of optical absorption spectra for the in-plane polarization as a function of wavelength, for aforementioned nanosheets and graphene [77,78] (black line) in the visible range (370–700 nm) of light.

We finally discuss the optical conductivity of these 2D systems. The real part of the optical conductivity is related to the Im $\varepsilon_{\alpha\beta}(\omega)$, and is given by [58];

$$\mathrm{Re}\,\sigma_{\alpha\beta}(\omega) = \frac{\omega}{4\pi}\mathrm{Im}\,\varepsilon_{\alpha\beta}(\omega) \quad (6)$$

The real part of the optical conductivity for the in-plane and out-of-plane polarizations are shown in Fig. 7. In order to provide a better vision, in this illustration we also included the corresponding results for the graphene monolayer. It is observable that the optical conductivities for the $C_7N_6$ monolayer along the both polarization directions start with a gap, originated from its semiconducting nature. In contrast, the optical conductivity for $C_9N_4$ and $C_{10}N_3$ monolayers yields singularity at zero frequency for the in-plane polarization, which is the signature of metallic materials. The first prominent optical conductivity peak along the in-plane polarization for $C_7N_6$ monolayer occurs at 3.15 eV. Our results indicate that the main peaks in the optical conductivities for the $C_7N_6$, $C_9N_4$ and $C_{10}N_3$ monolayers occur at energy ranges between 14.50-15.50 eV. The main peaks of Re $\sigma_{\alpha\beta}(\omega)$ of $C_7N_6$, $C_9N_4$ and $C_{10}N_3$ monolayers for the out-of-plane polarization locate at energy levels between 10.00 and 12.50 eV. Notably, the optical conductivities for the $C_7N_6$ and $C_9N_4$ nanosheets in the 370–550 nm range and for $C_{10}N_3$ structure in the 550-700 nm of light are higher than that of the graphene. These enhancements in the optical conductivities further highlight the desirable performances of these novel nanosheets for the application in optical nanodevices.

## 4. Summary

In summary, we introduced three novel carbon-nitride nanosheets, with porous atomic lattices and chemical formulas of $C_7N_6$, $C_9N_4$ and $C_{10}N_3$. These novel nanosheets include the cores made of three pentagons that are connected by single N atoms. Predicted nanomaterials were found to be dynamically and energetically stable, however only $C_7N_6$ and $C_9N_4$ nanosheets were estimated to be synthesizable as the $C_{10}N_3$ suffers from the thermal instability. Our theoretical results reveal that $C_7N_6$ and $C_9N_4$ nanosheets can exhibit high elastic modulus of 212 and 202 N/m and maximum tensile strengths of 14.1 and 22.4 N/m, respectively. As an astonishing finding, while $C_7N_6$ monolayer exhibits direct band-gap semiconducting electronic character with a 2.25 eV gap on the basis of HSE06 method, $C_9N_4$ was confirmed to be a metal, which is anomalous for the carbon-nitride 2D systems, which are mostly semiconductors. Optical calculations reveal that $C_7N_6$ monolayer can absorb the visible light, very promising for the applications in optoelectronics and nanoelectronics. Notably, $C_9N_4$ monolayer yields a semiconducting character for the optical spectra along the out-of-plane polarization. Our results confirm that the absorption coefficient and optical conductivity of predicted carbon-nitride 2D materials in the visible range of light are larger than those of the graphene, and are desirable for the employment in the optoelectronic nanodevices.

The contrasting electronic properties of $C_7N_6$ and $C_9N_4$ and only around 1% mismatch in their atomic lattices, propose that their lateral or planar heterostructures may show surprising electronic and optical features. We remind that $C_7N_6$ and $C_9N_4$ both include porous atomic lattices, which provide optimal conditions to reach active sites in these nanomaterials. On this basis and likely to other carbon-nitride 2D materials, $C_7N_6$ and $C_9N_4$ nanomembranes may also exhibit highly desirable properties for the application in some critical technologies, like: catalysis, photocatalysis and more importantly for the energy conversion/storage systems such as; rechargeable batteries, photovoltaic or hydrogen storage. According to these points and by also taking into consideration that both of $C_7N_6$ and $C_9N_4$ nanosheets are able to show remarkable mechanical rigidity as well as the dynamical and thermal stability, we are therefore hopeful that the findings provided by this study may serve as motivations for the experimental realization of these nanomaterials and also guide and stimulate further theoretical studies.

**Acknowledgment**

B. M. and X. Z. appreciate the funding by the Deutsche Forschungsgemeinschaft (DFG, German Research Foundation) under Germany's Excellence Strategy within the Cluster of Excellence PhoenixD (EXC 2122, Project ID 390833453). T. R. acknowledges the financial support by European Research Council for the COMBAT project (Grant number 615132).**Data availability**

The following are the supplementary data to this article:

**References**

[1]  K.S. Novoselov, A.K. Geim, S. V Morozov, D. Jiang, Y. Zhang, S. V Dubonos, I. V Grigorieva, A.A. Firsov, Electric field effect in atomically thin carbon films., Science. 306 (2004) 666–9. doi:10.1126/science.1102896.

[2]  A.K. Geim, K.S. Novoselov, The rise of graphene, Nat. Mater. 6 (2007) 183–191. doi:10.1038/nmat1849.

[3]  C. Lee, X. Wei, J.W. Kysar, J. Hone, Measurement of the Elastic Properties and Intrinsic Strength of Monolayer Graphene, Science (80-. ). 321 (2008) 385–388. doi:10.1126/science.1157996.

[4]  A.A. Balandin, S. Ghosh, W. Bao, I. Calizo, D. Teweldebrhan, F. Miao, C.N. Lau, Superior thermal conductivity of single-layer graphene, Nano Lett. 8 (2008) 902–907. doi:10.1021/nl0731872.

[5]  M. Yankowitz, S. Chen, H. Polshyn, Y. Zhang, K. Watanabe, T. Taniguchi, D. Graf, A.F. Young, C.R. Dean, Tuning superconductivity in twisted bilayer graphene, Science (80-. ). (2019). doi:10.1126/science.aav1910.

[6]  D.A. Bandurin, A. V. Tyurnina, G.L. Yu, A. Mishchenko, V. Zólyomi, S. V. Morozov, R.K. Kumar, R. V. Gorbachev, Z.R. Kudrynskyi, S. Pezzini, Z.D. Kovalyuk, U. Zeitler, K.S. Novoselov, A. Patanè, L. Eaves, I. V. Grigorieva, V.I. Fal'ko, A.K. Geim, Y. Cao, High electron mobility, quantum Hall effect and anomalous optical response in atomically thin InSe, Nat. Nanotechnol. (2016) 1–18. doi:10.1038/nnano.2016.242.

[7]  A.H.. Castro Neto, N.M.R.. Peres, K.S.. Novoselov, A.K.. Geim, F. Guinea, The electronic properties of graphene, Rev. Mod. Phys. 81 (2009) 109–162. doi:10.1103/RevModPhys.81.109.

[8]  M. Liu, X. Yin, E. Ulin-Avila, B. Geng, T. Zentgraf, L. Ju, F. Wang, X. Zhang, A graphene-based broadband optical modulator, Nature. 474 (2011) 64–67. doi:10.1038/nature10067.

[9]  F. Withers, M. Dubois, A.K. Savchenko, Electron properties of fluorinated single-layer graphene transistors, Phys. Rev. B - Condens. Matter Mater. Phys. 82 (2010). doi:10.1103/PhysRevB.82.073403.

[10] T.B. Martins, R.H. Miwa, A.J.R. Da Silva, A. Fazzio, Electronic and transport15

# Supporting Information

# Prediction of $C_7N_6$ and $C_9N_4$: Stable and strong porous carbon-nitride nanosheets with attractive electronic and optical properties

Bohayra Mortazavi[*a], Masoud Shahrokhi[b], Alexander V Shapeev[c], Timon Rabczuk[d] and Xiaoying Zhuang[a]

[a]*Institute of Continuum Mechanics, Leibniz Universität Hannover, Appelstraße 11, 30157 Hannover, Germany.*
[b]*Department of Physics, Faculty of Science, Razi University, Kermanshah, Iran.*
[c]*Skolkovo Institute of Science and Technology, Skolkovo Innovation Center, Nobel St. 3, Moscow 143026, Russia.*
[d]*College of Civil Engineering, Department of Geotechnical Engineering, Tongji University, Shanghai, China.*

*E-mail: bohayra.mortazavi@gmail.com

1. Atomic structures of constructed monolayers unit-cells in VASP POSCAR.

2- AIMD results for the thermal stability.

3- Failure mechanism upon the uniaxial loading.

4- HSE06 results for the total electronic density of states.

1. Atomic structures of constructed monolayers unit-cells in VASP POSCAR.

1.1-C7N6
  1.00000000000000
    6.7943948564804693    0.0000000000000000    0.0000000000000000
    3.3971974282402351    5.8841185491832784    0.0000000000000000
    0.0000000000000000    0.0000000000000000   15.0000000000000000
   C   N
   7   6
Direct
  0.9998100576047193  0.0002509396813224  0.5000000000000000
  0.9331640163519950  0.7159702217766863  0.5000000000000000
  0.9331204068991781  0.3511955830278453  0.5000000000000000
  0.7155979589857679  0.9335029314294445  0.5000000000000000
  0.7155586812735208  0.3512017827147176  0.5000000000000000
  0.3507638336030467  0.9335002995972488  0.5000000000000000
  0.3507380370779813  0.7159747027046279  0.5000000000000000
  0.1110737155275103  0.1115015469972604  0.5000000000000000
  0.1110716970172021  0.7777302000807538  0.5000000000000000
  0.9928516147958462  0.5037223459642206  0.5000000000000000
  0.7772919263856721  0.1115054214509428  0.5000000000000000
  0.5034145543584359  0.9930456207717029  0.5000000000000000
  0.5032887640754495  0.5037339139401453  0.5000000000000000

1.2-C9N4
  1.00000000000000
    6.8747763944394347    0.0000000000000000    0.0000000000000000
    3.4373881972197169    5.9537310029223516    0.0000000000000000
    0.0000000000000000    0.0000000000000000   15.0000000000000000
   C   N
   9   4
Direct
  0.9393400458179499  0.7131257480603921  0.5000000000000000
  0.9393001191584816  0.3477474643952050  0.5000000000000000
  0.7127859420819433  0.9397274414772028  0.5000000000000000
  0.7128273967788488  0.3477398726273222  0.5000000000000000
  0.3473822033077525  0.9397377476090796  0.5000000000000000
  0.3474029143193675  0.7131344478612078  0.5000000000000000
  0.1122806251630522  0.1126744582075219  0.5000000000000000
  0.1122974051619892  0.7752802886217438  0.5000000000000000
  0.7749246449094613  0.1126685991057670  0.5000000000000000
  0.9959374896874280  0.5021250988652426  0.5000000000000000
  0.5017923976542491  0.9963276492876858  0.5000000000000000
  0.5018066045797767  0.5021285977240453  0.5000000000000000
  0.9998464096288657  0.0001947787100747  0.5000000000000000

1.3-C10N3
  1.00000000000000
    6.9479592820170710    0.0000000000000000    0.0000000000000000
    3.4739796410085360    6.0171092426870594    0.0000000000000000
    0.0000000000000000    0.0000000000000000   15.0000000000000000
  C  N
  10  3
Direct
 0.9998618242113650  0.0001915034646203  0.5000000000000000
 0.9421341773494589  0.7110490657941355  0.5000000000000000
 0.9421032766220279  0.3470382183706917  0.5000000000000000
 0.7107023464341822  0.9425717864329997  0.5000000000000000
 0.7107678193893250  0.3469847987877088  0.5000000000000000
 0.3466836888252871  0.9425567476066306  0.5000000000000000
 0.3467199055403896  0.7109873132782667  0.5000000000000000
 0.1136681523335525  0.1141695712157755  0.5000000000000000
 0.1136820138782042  0.7724426820622625  0.5000000000000000
 0.7719589577274917  0.1141792480925332  0.5000000000000000
 0.9832906640111574  0.5084824128072398  0.5000000000000000
 0.5080726875267416  0.9837057033943637  0.5000000000000000
 0.5080998301508259  0.5084764676927023  0.5000000000000000

2- AIMD results for the thermal stability.

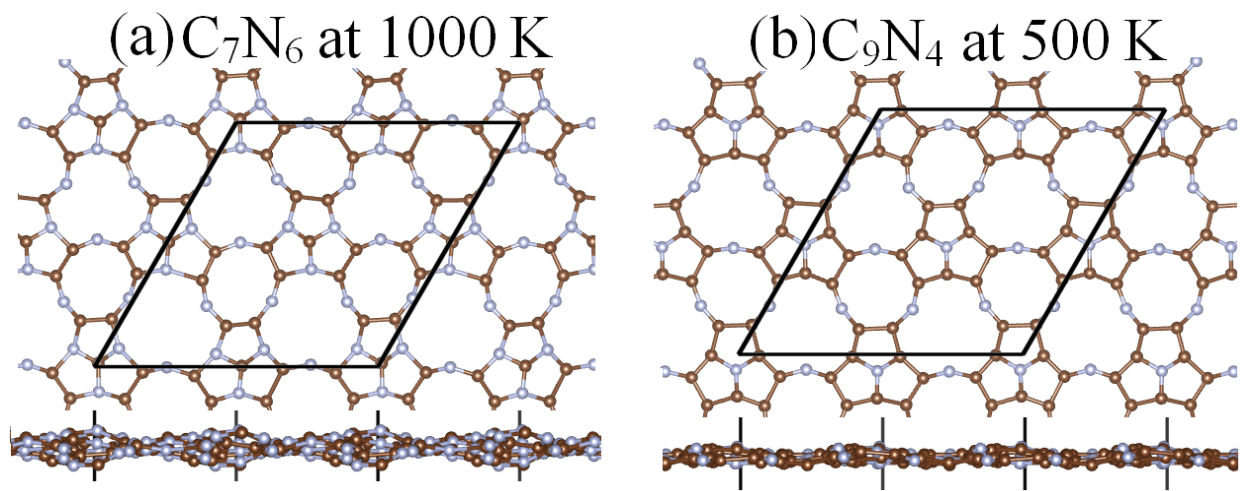

Fig. S1, Top and side views of $C_7N_6$ and $C_9N_4$ monolayers after the AIMD simulations for 20 ps.

3- Failure mechanism upon the uniaxial loading.

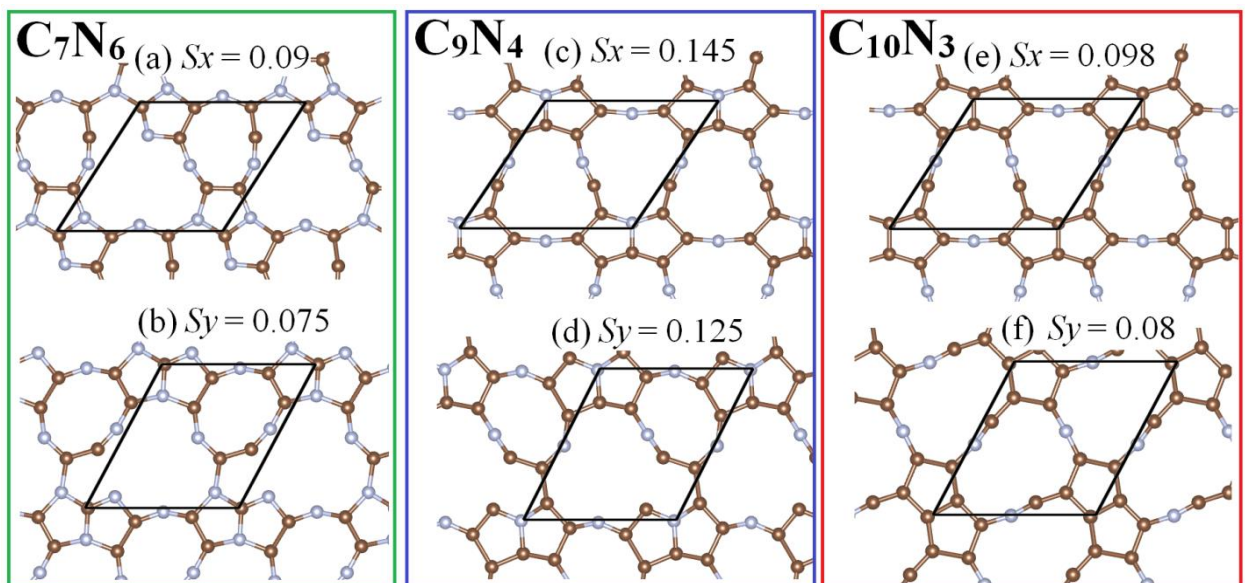

Fig. S2, Top views of the single-layer $C_7N_6$, $C_9N_4$ and $C_{10}N_3$ at strain levels shortly after the ultimate tensile strength point. $S_x$ and $S_y$ depict the strain levels for the uniaxial loading along the $x$ than $y$ directions, respectively

4- HSE06 results for the total electronic density of states.

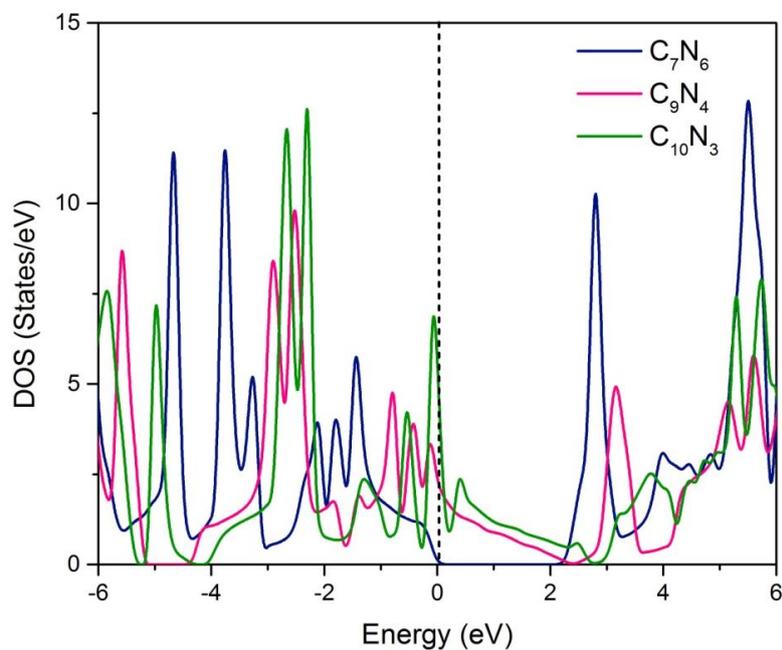

Fig. S3, Total electronic density of states predicted by the HSE06 functional. The Fermi energy is aligned to zero.

26